\documentclass[twoside,dvips]{elsart}
\usepackage{epsfig,amsmath}
\def\beq{\begin{equation}}
\def\eeq{\end{equation}}

\begin{document}
\begin{frontmatter}

\vspace{-15pt}
\title{ Study of RPC gas mixtures for the ARGO-YBJ experiment}
\vspace{-20pt}
\vspace{-8pt}
\author{B. Bartoli}, \author{R. Buonomo}, 
\author{E. Calloni}, \author{S. Catalanotti}, 
\author{B. D'Ettorre Piazzoli}, \author{G. Di Sciascio}, 
\author{M. Iacovacci}

\address{INFN and Dipartimento di Fisica dell'Universit\`a di Napoli, Italy}

\begin{abstract}
The ARGO-YBJ experiment consists of a RPC carpet to be operated at the 
Yangbajing laboratory (Tibet, P.R. China), 4300 m a.s.l., and devoted to the 
detection of showers initiated by photon primaries in the energy range 
100 GeV - 20 TeV. 
The measurement technique, namely the timing on the shower front with a 
few tens of particles, requires RPC operation with 1 ns time resolution, 
low strip multiplicity, high efficiency and low single counting rate. 
We have tested RPCs with many gas mixtures, at sea level, in order to 
optimize these parameters. The results of this study are reported.
\end{abstract}

\end{frontmatter}

\vspace{-30pt}

\section{Introduction}
\vspace{-18pt}

The ARGO-YBJ experiment \cite{argop} is under way over the next few years 
at the Yangbajing High Altitude Cosmic Ray Laboratory (4300 m a.s.l., 
606 $g/cm^2$), 90 km North to Lhasa (Tibet, P.R. China).
The aim of the ARGO-YBJ experiment is the study of fundamental issues in 
Cosmic Ray and Astroparticle Physics including $\gamma$-ray astronomy, 
GRBs physics at 100 GeV threshold energy and the measurement of the 
$\overline{p}/p$ at TeV energies.
The apparatus consists of a full coverage detector of dimension 
$71\times 74$ $m^2$ realized with a single layer of Resistive Plate 
Counters (RPCs). A guard ring partially (about 50 $\%$) instrumented 
with RPCs, surrounds the central detector, up to $100\times 100$ $m^2$; 
it improves the apparatus performance by enlarging the fiducial area for the 
detection of showers with the core outside the full coverage carpet.

A lead converter 0.5 cm thick will cover uniformly the RPC plane in order 
to increase the number of charged particles by conversion of shower photons 
and to reduce the time spread of the shower particles. 
The measurement technique, namely the timing on the shower front with a few 
tens of particles, requires RPC operation with 1 ns time resolution, low 
strip multiplicity for good energy estimation at low energies, high 
efficiency and low single counting rate to trigger efficiently at low 
multiplicity. 

Keeping in mind all these needs and the low operating pressure (about 600 
mbar) at Yangbajing we started to investigate different gas mixtures, at 
sea level, in order to optimize the detector performance. In fact previous 
studies have shown that the performance of the detector may be heavily 
affected by the reduced pressure \cite{natali}.
Three gas components were used: Argon, iso-Butane C4H10 and TetraFluoroEthane 
C2H2F4 that will be indicated in the following as Ar, i-But and TFE 
respectively. 

The set-up used for this study consists of a small telescope of 4 RPCs 
$50\times 50$ $cm^2$ area with 16 pick-up strips 3 cm wide connected to the 
front-end electronics board .
The front-end circuit contains 16 discriminators, with about 70 mV voltage 
threshold, and provides a FAST-OR signal with the same input-to-output 
delay (10 ns) for all the channels.
The 4 RPCs  were overlapped one on the other, 3 out of them were used to 
define a cosmic ray beam by means of a triple coincidence of their FAST-OR 
signals, the fourth one was used as test RPC.
The three RPCs providing the trigger were operated with a gas mixture of 
$60\%$ Ar, $37\%$ i-But and $3\%$ TFE.
At any trigger occurrence the time provided by  the test RPC was read  by 
means of  a LECROY TDC of 0.25 ns time bin, operated in common START mode; 
the number of fired strips was read by means of a CAEN module C187. 
The single counting rate was read by a CAEN scaler C243. 

\vspace{-25pt}
\section{RPC performance}
\vspace{-18pt}

The RPCs were operated in streamer mode, as foreseen for the experiment, at 
the ARGO laboratory of the Physics Department of the Naples University. 

\begin{figure}[htb]
\hspace{-2cm}
\vfill \begin{minipage}{.47\linewidth}
\begin{center}
\mbox{\epsfig{file=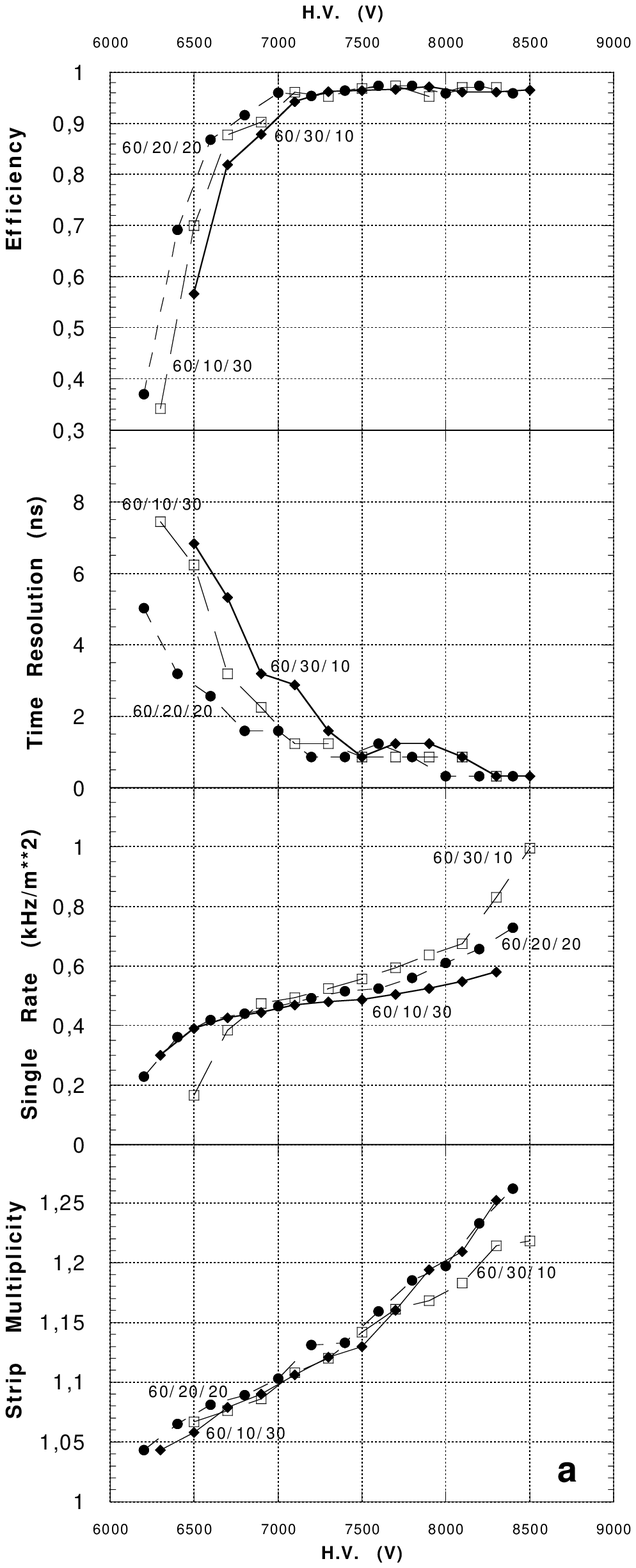,height=19.cm,width=13.cm}}
\end{center}
\end{minipage}\hfill
\hspace{-2cm}
\begin{minipage}{.47\linewidth}
\begin{center}
\mbox{\epsfig{file=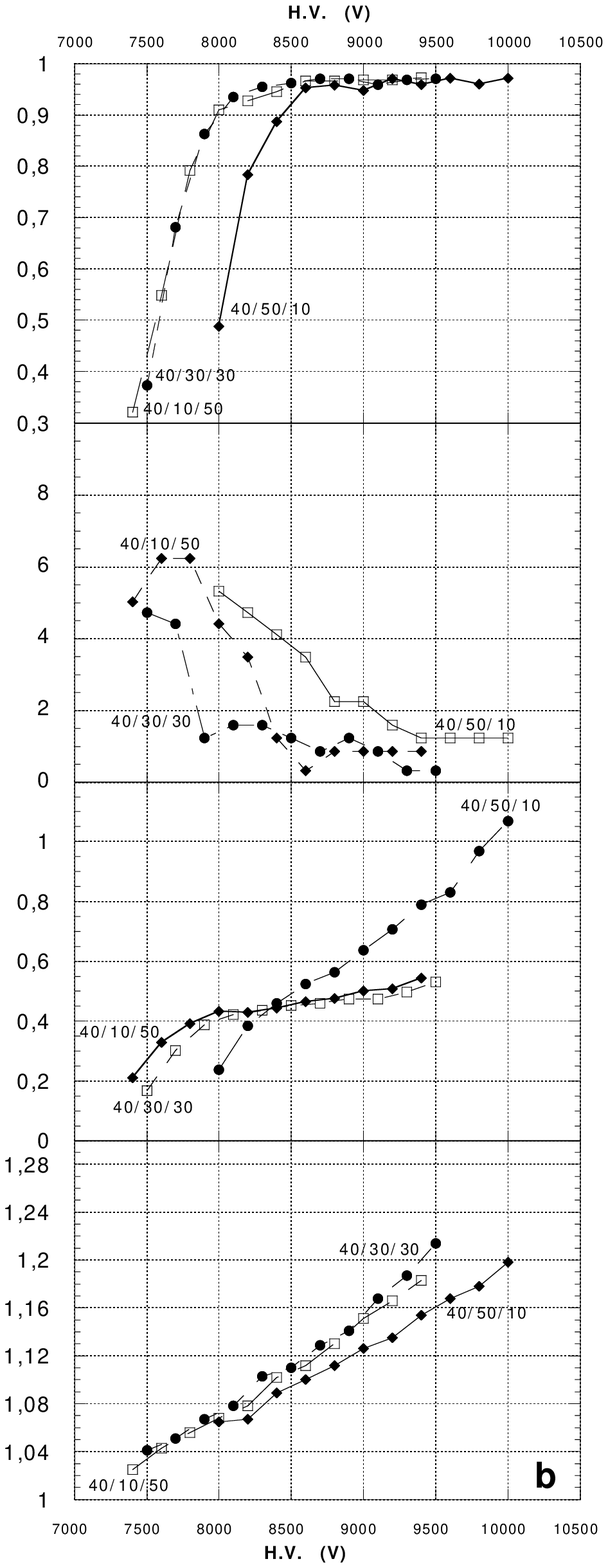,height=19.cm,width=13.cm}}
\end{center}
\end{minipage}\hfill
\caption{Efficiency, time resolution, single rate and mean strip 
multiplicity for gas mixtures with a) $60\%$ of Ar; b) $40\%$ of Ar.}
\label{fig1}
\end{figure}

Many gas mixtures have been tested, including mixtures with a high 
percentage of Ar ($40\%$ - $60\%$) which represent a reference point for 
the performance of the detector. 
The results are shown in Fig. \ref{fig1} where in the first column a) are 
reported the three mixture with Ar kept at $60\%$; the numbers associated 
to each line like 60/30/10 refer respectively to Ar/i-But/TFE. 
It can be seen that changing the relative percentage of the two quenching 
components, i-But and TFE, does not result in any strong or evident 
difference of the detector performance. In fact the operating voltage does 
not chance substantially, neither the efficiency; the time resolution 
stabilizes around 1 ns and the strip multiplicity ranges in 1.1 - 1.3 
depending on the applied voltage. 
This is confirmed by the set of measurements done with Ar kept at $40\%$ 
(column b). Only the mixture 40/50/10 shows some difference, but it is likely 
that there was some uncontrolled change or problem during the data taking. 

\begin{figure}[htb]
\vfill \begin{minipage}{.47\linewidth}
\begin{center}
\mbox{\epsfig{file=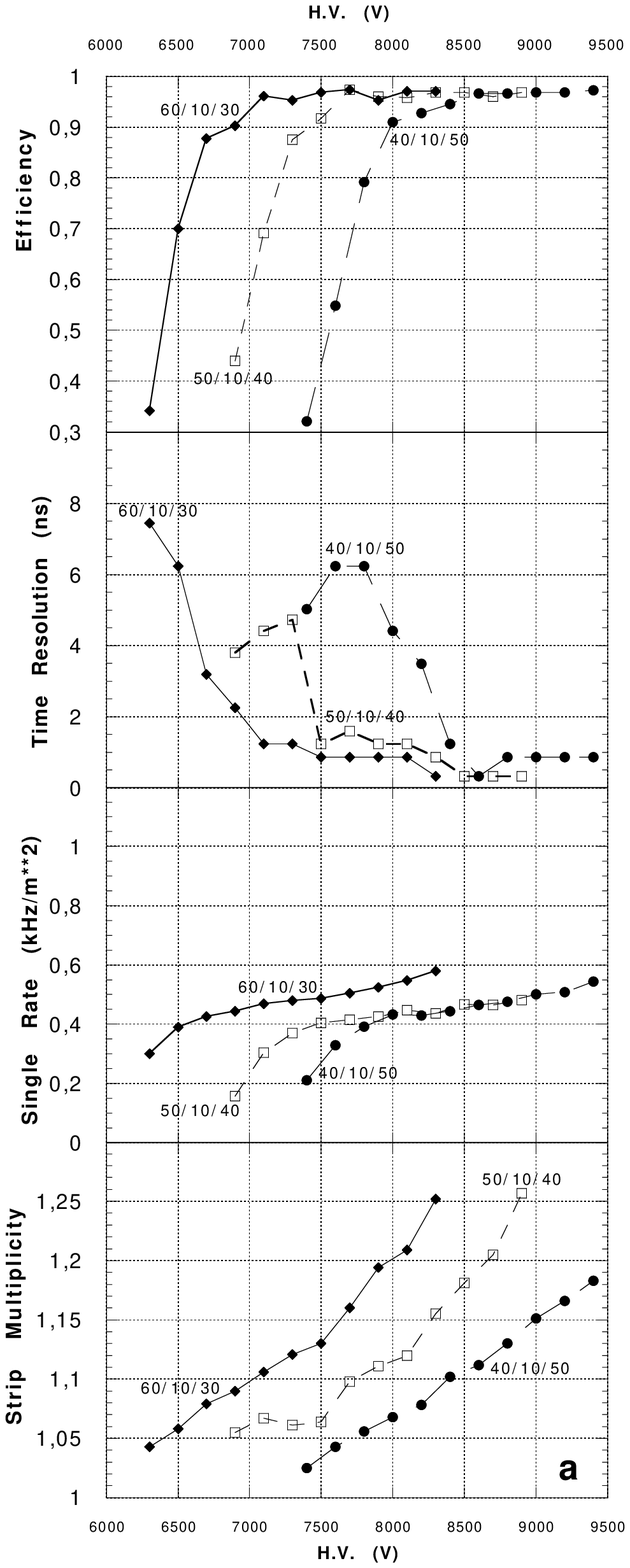,height=19.cm,width=13.cm}}
\end{center}
\end{minipage}\hfill
\hspace{-2cm}
\begin{minipage}{.47\linewidth}
\begin{center}
\mbox{\epsfig{file=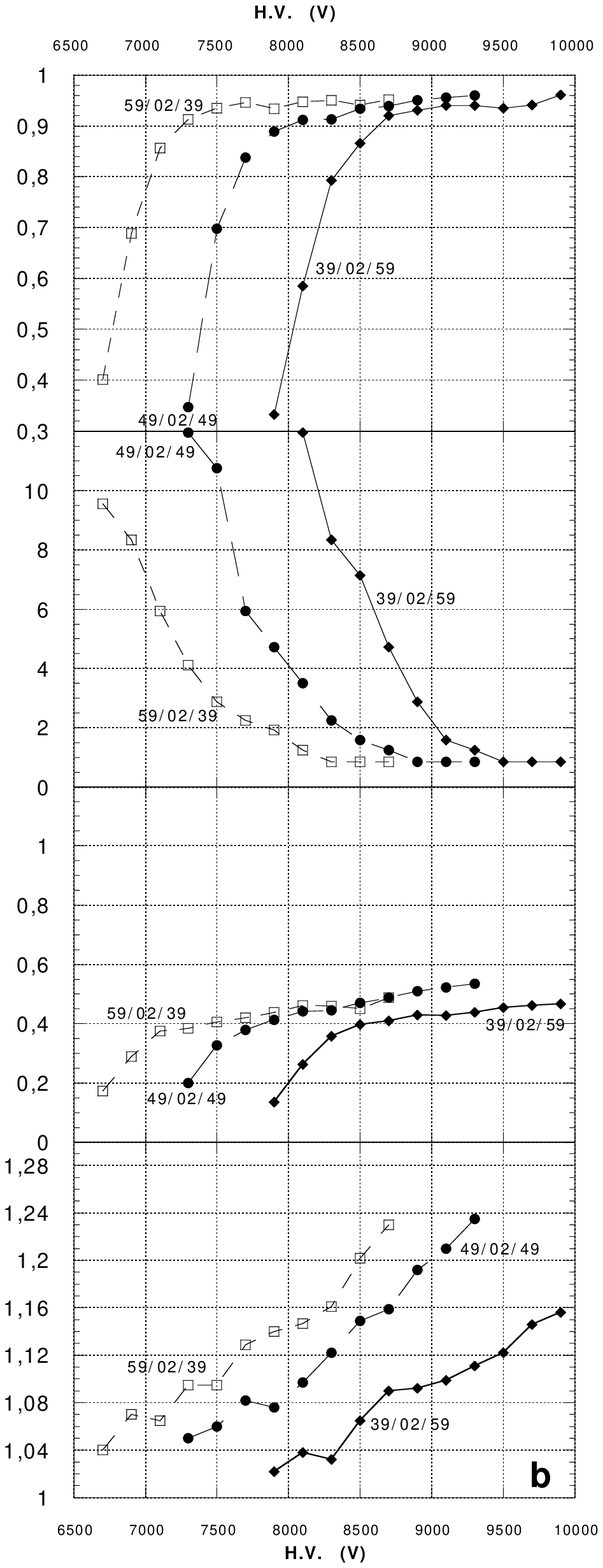,height=19.cm,width=13.cm}}
\end{center}
\end{minipage}\hfill
\caption{Efficiency, time resolution, single rate and mean strip multiplicity 
for gas mixtures with a) $10\%$ of i-But; b) $2\%$ of i-But.}
\label{fig2}
\end{figure}

In order to check gas mixtures with i-But content around and below the 
flammability threshold ($10\%$) we have performed measurements with mixtures 
where the i-But fraction was kept at $10\%$ in one set and $2\%$ in another 
set. The results are reported in  column a) and b) of  Fig. \ref{fig2}.
Both samples of measurements show that the Ar percentage plays an important 
role in defining the operating voltage (about 800 V above the knee-voltage
$V_{knee}$) and influences the strip multiplicity. 
The mean value of the strip multiplicity remains anyway below 1.3; also the 
single rate changes according to the Ar percentage and this can be 
understood  as we work at fixed discrimination threshold. 
Efficiency and time resolution are not affected. This means that i-But, 
traditionally used to "quench" the discharge by asbsorption of ultraviolet 
photons, can be substituted almost completely by TFE which is electronegative 
and captures electrons of the plasma. 

\begin{figure}
  \begin{center}
    \mbox{\epsfig{file=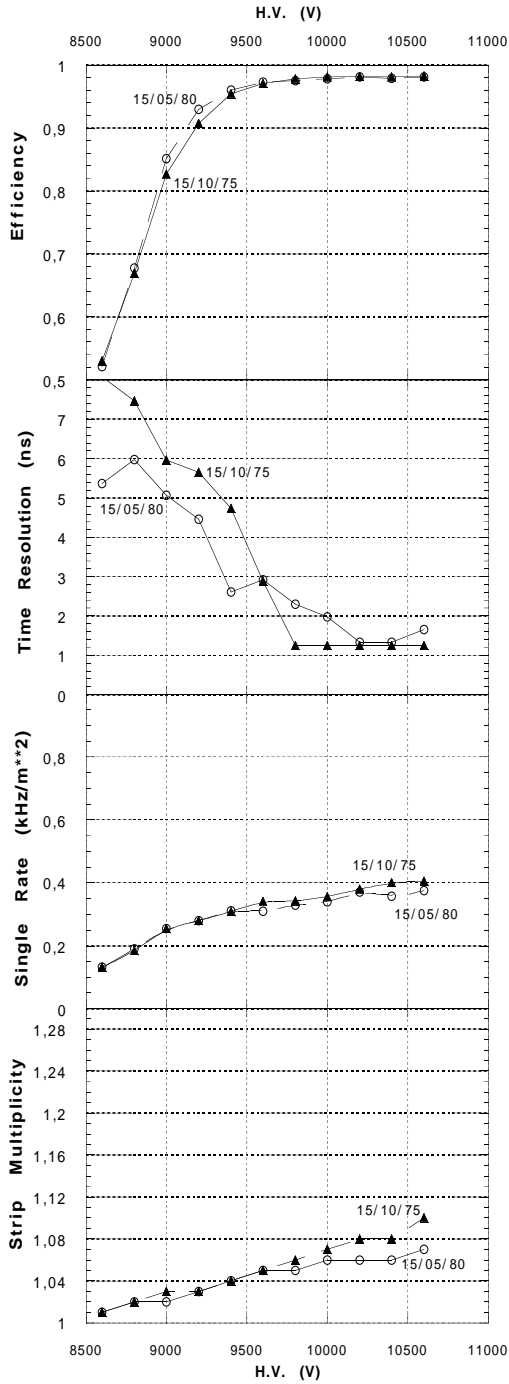,height=19.cm,width=13.cm}}
    \caption{Efficiency, time resolution, single rate and mean strip 
multiplicity for the gas mixtures Ar/i-But/TFE = 15/10/75 and 
Ar/i-But/TFE = 15/5/80. }
    \label{fig3}
  \end{center}
\end{figure}

A solution to the problem of low pressure is increasing the density of the 
gas mixture. An increase of TFE concentration at expenses of the Ar 
concentration should therefore increase the primary ionization thus 
compensating for the $40\%$ reduction caused by the lower gas target 
pressure (600 $mbar$) and reduces the afterpulse probability. 
We have tested two mixtures with high content of TFE, namely  Ar/i-But/TFE : 
15/10/75 and 15/5/80; the second one is a non flammable mixture. 
The experimental results ( Fig. \ref{fig3}) do not exhibit evident 
differences between them. 
The reduction of the Argon concentration in favour 
of TFE results in a clear increase of the operating voltage as expected from 
the large quenching action of TFE. The  efficiency, $98\%$, is comparable 
to the values obtained with the mixtures having low percentage of TFE; 
might even be higher but this, given the experimental fluctuations, cannot be 
stated with certainty. Moreover, the time resolution doesn't show any 
worsening, the single rate keeps below 400 $Hz/m^2$ and also the strip 
multiplicity stays below 1.1. The mixture 15/10/75 has been already used 
successfully at Yangbajing in the ARGO test \cite{bacci99a} \cite{bacci99b}.

\vspace{-25pt}
\section{Conclusions}
\vspace{-18pt}

We have studied the performance of RPCs with gas mixtures made of Ar, i-But 
and TFE. We checked mixtures with a high percentage of Ar, mixtures with 
low content of  i-But, and finally mixtures with low and high percentage 
of TFE. For all of them, apart from those with a high percentage of TFE, 
we can summarize as follows: 1) the efficiency is $95\div 97\%$; 
2) the voltage where the efficiency plateau starts and the working voltage 
(tipically $V_{knee}$ + 800 V) depends strongly on the Ar fraction, showing 
an increase of about 500 V per every $10\%$ Ar reduction); 3) the single 
rate shows a plateau in the frequency range 400 $Hz/m^2$ - 600 $Hz/m^2$; 
4) the time resolution at working voltage is tipically $1.0\div 1.3$ $ns$; 
5) the mean value of the strip multiplicity at the working voltage is 
$1.15\div 1.25$ depending on the Ar percentage. 

The mixtures with higher TFE content result in sligthly better performance, 
giving $98\%$ efficiency, lower strip multiplicity and single rate. 
We couldn't find any parameter able to discriminate the different mixtures. 
Preliminary measurements show that in mixtures with higher TFE fraction a 
smaller charge per track is developed. Tests are going on to investigate the 
analog read-out of the detector operated in streamer mode.


\end{document}